\documentclass[conference]{IEEEtran}
\IEEEoverridecommandlockouts
\usepackage{amsmath,amssymb,amsfonts}
\usepackage{algorithmic}
\usepackage{graphicx}
\usepackage{textcomp}
\usepackage{xcolor}
\def\BibTeX{{\rm B\kern-.05em{\sc i\kern-.025em b}\kern-.08em
    T\kern-.1667em\lower.7ex\hbox{E}\kern-.125emX}}
\usepackage[]{mdframed}
\usepackage{booktabs}
\usepackage{multirow}
\usepackage[justification=centering]{caption}
\usepackage{adjustbox}
\usepackage{longtable}
\usepackage{array}
\usepackage{threeparttable}
\usepackage{adjustbox}
\usepackage{lipsum}
\usepackage{url}
\usepackage{listings}
\usepackage{xcolor}
\usepackage{geometry}
\usepackage[hidelinks]{hyperref}
\usepackage{tabularx}
\usepackage[backend=biber,style=ieee,sorting=none]{biblatex}
\usepackage{todonotes}
\lstdefinelanguage{Solidity}{
    keywords={contract, function, if, else, for, while, return, uint, int, string, bool, mapping, address, public, private, pragma, solidity, import, assertEq},
    comment=[l]{//},
    morecomment=[s]{/*}{*/},
    commentstyle=\color{gray}\itshape,
    basicstyle=\ttfamily,
    morestring=[b]",
    morestring=[b]',
    morestring=[b]`
}

\lstdefinestyle{solidityStyle}{
    language=Solidity,
    basicstyle=\ttfamily\footnotesize,
    backgroundcolor=\color{white},
    showspaces=false,
    showstringspaces=false,
    showtabs=false,
    frame=single,
    tabsize=2,
    captionpos=b,
    breaklines=true,
    breakatwhitespace=false,
    escapeinside={\%*}{*)},
    xleftmargin=5pt,  % Adjust the left margin
    xrightmargin=5pt  % Adjust the right margin
}

\lstdefinestyle{englishStyle}{
    basicstyle=\ttfamily\footnotesize,
    backgroundcolor=\color{white},
    showspaces=false,
    showstringspaces=false,
    showtabs=false,
    frame=single,
    tabsize=2,
    captionpos=b,
    breaklines=true,
    breakatwhitespace=false,
    escapeinside={\%*}{*)},
    breakindent=0pt,  % No indentation before line breaks
    xleftmargin=5pt,  % Adjust the left margin
    xrightmargin=5pt  % Adjust the right margin
}
\makeatletter
\newcommand{\linebreakand}{%
\end{@IEEEauthorhalign}
\hfill\mbox{}\par
\mbox{}\hfill\begin{@IEEEauthorhalign}
}
\makeatother

\begin{document}

\title{Towards LLM-assisted High-Quality Property Generation for Solidity Smart Contracts\\

}

\author{
	\IEEEauthorblockN{Muhammad Wahid}
	\IEEEauthorblockA{FAST-NUCES, Pakistan\\f200175@cfd.nu.edu.pk}
	\and
	\IEEEauthorblockN{Shahzaib Khan}
	\IEEEauthorblockA{FAST-NUCES, Pakistan\\f201079@cfd.nu.edu.pk}
	\and
	\IEEEauthorblockN{Mashhood Ali}
	\IEEEauthorblockA{FAST-NUCES, Pakistan\\f200219@cfd.nu.edu.pk}
	\linebreakand % This forces a new line for the next set of authors
	\IEEEauthorblockN{Muhammad Hassan}
	\IEEEauthorblockA{Truscova GmbH, Germany\\hassan@truscova.com}
	\and
	\IEEEauthorblockN{Muhammad Naiman Jalil}
	\IEEEauthorblockA{United Arab Emirates University, UAE\\muhammad.jalil@uaeu.ac.ae}
	\and
	\IEEEauthorblockN{Affan Rauf}
	\IEEEauthorblockA{FAST-NUCES, Pakistan\\affan.rauf@nu.edu.pk}
}
\maketitle
\thispagestyle{plain}

\begin{abstract}
The immutable nature of smart contracts makes it challenging to fix and patch bugs once they are deployed to a blockchain.
This implies that security vulnerabilities may be exposed to possible exploitation for a longer period, necessitating comprehensive pre-deployment testing. 
Property-based testing combined with fuzzing has proven itself as a promising technique for uncovering vulnerabilities. 
%Unlike traditional example-based testing, property-based testing aims to ensure that a system consistently adheres to general properties across all possible states and inputs. 
Traditionally, system properties are written by human experts, which is time-consuming and consequently expensive. With the recent advancement in Large Language Models (LLMs) and their ability to `understand' natural language and code semantics, it may be possible to generate effective properties. This study, 
%we propose an approach that 
leverages state-of-the-art LLMs to generate high-quality properties for Solidity-based smart contracts. 
%assist property-based testing by automatically . Once the properties are generated, we apply fuzzing to them. 
We measure the quality of the generated properties using mutation testing. Our results show that LLMs have the potential to generate high-quality properties that are close to those written by human experts. We extensively evaluate LLMs using various prompting techniques (e.g., zero shot, few shot, and prompt chaining). Overall, we find that Gemini Pro 1.5, when combined with prompt chaining, achieves the highest average mutation score of 25.99\% among all studied configurations, closely approaching the human written benchmark of 31.75\%. However, our per contract analysis reveals notable variance, particularly for the LibBit contract, where Gemini Pro 1.5 under prompt chaining achieves a mutation score of 74.34\%, which is on par with human written properties (74.83\%). This highlights that while average performance is informative, individual contract level results demonstrate that LLMs can, in some cases, match expert level property generation. 
\end{abstract}

\renewcommand{\footnoterule}{
    \vspace*{2pt}
    \noindent\rule{0.4\linewidth}{0.4pt}
    \vspace*{2pt}
}

%\footnotetext[1]{ Shahzaib Khan \& Muhammad Wahid are the co-first authors.}
%\footnotetext[2]{ Affan Rauf is the corresponding author.}

\section{Introduction}
Blockchain technology has recently emerged as a powerful innovation that is transforming centralized systems into decentralized networks. It has revolutionized various sectors, including education, healthcare, and, most significantly, the finance sector. At the heart of this technology are smart contracts, self-executing immutable agreements that enforce and execute business rules without any intermediaries in a decentralized manner, ensuring security and transparency. 
Smart contracts are prone to various types of vulnerabilities, such as integer overflow~\cite{b1} and reentrancy attacks~\cite{b2, b3}. In recent years, it has faced significant security breaches, including the infamous DAO attack, which resulted in the loss of 3.6 million ethers~\cite{b4}. Many of these security breaches occur due to loopholes in smart contracts. To address these issues early on in the design phase, thorough testing of smart contracts is essential. 

Traditional testing methods, such as example-based testing, have long been employed to validate software functionality against predefined test values. However, as software systems grow increasingly complex and interconnected, the limitations of these traditional approaches become more apparent. Property-based testing (PBT)~\cite{b5} emerges as a powerful alternative. Unlike traditional testing methods that rely on a finite set of example inputs and expected outputs, PBT focuses on defining general properties that a program should satisfy for all inputs. By Fuzzing i.e., automatically running the system for randomly generated inputs, PBT can uncover unexpected behaviors and corner cases that are difficult to identify through manual test input design. For complex systems where input spaces are vast, PBT offers a scalable approach to ensure correctness.

One of the primary hurdles in PBT is the manual effort involved in defining system properties. Writing effective properties requires a deep understanding of the software's expected behavior, which can be time-consuming and complex. Additionally, the cost associated with this manual effort can be prohibitive, especially for large-scale projects or organizations with limited resources. These challenges have limited the broader adoption of PBT, as the benefits of thorough testing are weighed against the resources required to implement it effectively.

The recent advancement in Large Language Models (LLMs) presents an opportunity to reduce manual effort in writing properties. LLMs have demonstrated remarkable capabilities in understanding natural language descriptions and generating relevant source code. 
Moreover, LLMs have recently been used to generate property-based tests for various programming languages~\cite{b6},~\cite{b7},~\cite{b8} indicating their potential to assist property-based testing.
However, the effectiveness of LLMs in this context remains an open question.
In this paper, we 
%evaluate various LLMs on different prompting techniques 
address this key question i.e., Can LLMs generate properties for Solidity smart contracts that are of high-quality, i.e., as effective as those written by human experts?

%To address this question, 
In this study, we present Smart Property Generator that leverages state-of-the-art LLMs using various prompting techniques for generating test properties for solidity-based smart contracts. In particular, we employ zero-shot prompting, few-shot prompting, and prompt chaining on GPT-4-turbo, GPT-4o, and Gemini-pro-1.5 to guide them in generating test properties. Our approach utilizes the existing capabilities of LLMs without the need for additional training. 

We use Smart Property Generator to generate properties for real-world smart contracts available in the Solady project~\cite{b9}, a widely used Ethereum utility library with approximately 30,000 weekly downloads, as reported by ~\cite{solady-npm-stat}. Our focus is specifically on the smart contracts within Solady’s utils folder, which offers a collection of low-level components frequently employed across various decentralized applications. We then compile and fuzz these smart contracts using Foundry~\cite{b10}, a smart contract development framework. Fuzzing~\cite{b11} is a popular technique for identifying security vulnerabilities by testing them on a wide range of diverse inputs. We benchmark the quality of properties generated by the LLMs using mutation score i.e., the ratio of the number of killed mutants (modified program) to the total number of non-equivalent mutants. A higher mutation score indicates a more effective property in catching bugs. Moreover, we measure the effectiveness and efficiency of different LLMs for the generating properties using the percentage of executable and non-executable properties, and the inference time, respectively. Our findings show that Gemini-pro-1.5 combined with prompt chaining achieves a mutation score of 25.99\%, approaching close to the score of human-generated properties which was 31.75\%.

This paper makes the following contributions: 
\begin{itemize}
    \item It proposes a approach to generate high-quality properties for solidity-based smart contracts using three prompting techniques on state-of-the-art LLMs.
    \item It evaluates the effectiveness of LLM-generated properties and compares them with human-written properties using mutation testing to ensure their quality.
\end{itemize}

\section{LLM-Assisted Approach}

We have relied on the knowledge base of LLMs for this research without fine-tuning the LLMs for our specific use case. Our research aims to evaluate the LLMs’ capabilities for generating Solidity test properties. We evaluated different LLMs, including GPT-4Turbo, GPT-4o, and Gemini-1.5-pro. We assessed each LLM using different prompts to gain a better understanding of their effectiveness for generating properties. 

\subsection{Overview}
\begin{figure*}
  \includegraphics[width=\textwidth,height=4cm]{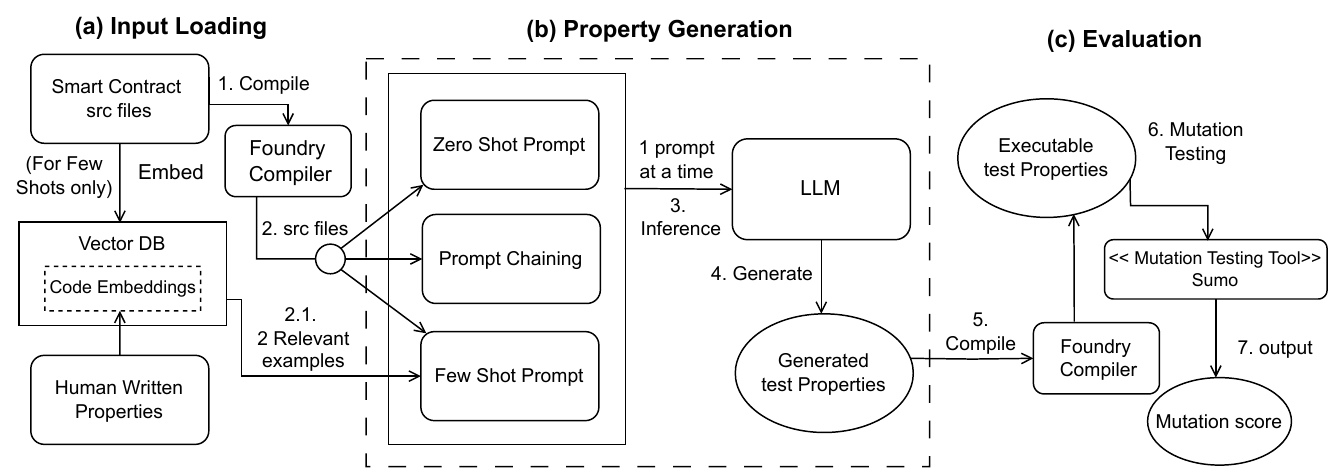}
  \caption{A high-level workflow of the LLM-assisted approach for generating high-quality properties for Solidity smart contracts.}
  \label{block-diagram}
\end{figure*}

Our LLM-assisted approach consists of two modules: input loading and property generation, as illustrated in Fig.~\ref{block-diagram}.
First, in the input loading phase, we take all source files for the smart contracts to provide the LLM with complete context on each functionality. We compile these files using the Foundry compiler. If we encounter any compilation issues, we provide an executable source file. Refer to Fig.~\ref{block-diagram}(a) for an illustration of this process.
Next, as shown in Fig.~\ref{block-diagram}(b), we proceed to the property generation module, which employs three different prompting techniques. We start by selecting a prompt and integrating the source files into the prompt template. For few-shot prompts, we also retrieve two relevant examples from the vector database and include them in the prompt template. After finalizing the prompt, we pass it to the LLM for inference, and the LLM generates the test properties for the smart contracts.
In the evaluation phase, depicted in Fig.\ref{block-diagram}(c), we select only those test properties from the generated set that are executable or can be made executable with minimal manual effort. We then perform mutation testing on these selected properties using Sumo\cite{b13}, an automated mutation testing tool compatible with Foundry\cite{b37}. After completing the mutation testing, we obtain the mutation score of the properties.

In the next sections, we explain the LLM-assisted approach in detail.

\subsection{Input Loading}
Before incorporating the input code into the prompt template, we first verify that all provided source files compiled successfully. This verification is managed by the input loader module, which precedes the properties generation module.
The input loader module operates in two phases. Initially, all source files are placed in the \verb|src| folder of the Foundry project, which includes smart contracts, libraries, interfaces and utils. Subsequently, we compile all the source files using the Foundry compiler. We proceed to the properties generation module only after confirming that all .sol files compile successfully. If any compilation issues arise, the user must provide executable code to resolve them.

\subsection{Properties Generation}
This module aims to generate Foundry test properties for Solidity smart contracts, covering as many edge cases as possible. For effective testing, the generated test properties must meet the following criteria:
\begin{itemize}
\item They must be executable.
\item They must adhere to the property-based testing syntax and standards provided by Foundry.
\item They must cover all possible edge cases.
\end{itemize}

The property generator module takes a Solidity smart contract file from the \verb|src| folder, along with any imported .sol files, and incorporates them into the prompt template. We then pass the prompt, which includes natural language instructions and the code, to the LLM. The LLM generates test properties for the specific source smart contract file. It is important to note that the output of LLM varies even on the same prompt. So we made five requests for each smart contract for each prompt used and selected the output having more executable properties. To determine if the generated properties are executable or not, we ran them using Foundry and then passed the selected output to Foundry to run the generated test properties for mutation testing. Following Foundry standards, we add the generated test properties to a t.sol file and place it in the \verb|test| folder of the Foundry project. Finally, we run the tests using Foundry. For more details on the hallucinations in the LLM's output, see Section~\ref{subsec:observations}.

We have employed various prompts to evaluate the performance of LLMs in generating properties for Solidity smart contracts. Details of each prompt are discussed in the following subsections.

\paragraph{Zero-Shot} The zero-shot prompt~\cite{b14} only contains natural language instructions for properties generation, as shown in Listing~\ref{lst:zeroShotPrompt} and specifies the output format, as demonstrated in Listing~\ref{lst:output_test_contract_format}. We do not provide any code as an example. In the output format, we only provide the structure of the test contract so that the LLM follows the Foundry testing standards and imports every time. It was repeatedly observed that without specifying the output format, the generated test contracts often missed necessary imports, leading to compilation errors.

% Listings 1,2
\begin{lstlisting}[style=englishStyle, caption={Zero Shot Prompt}, label=lst:zeroShotPrompt]
Write foundry test smart contract code in solidity, containing test property for each functionality of the smart contract, delimited by ####.
####
Give me the solidity code for each test property.
####
\end{lstlisting}

\begin{lstlisting}[style=solidityStyle,caption={Output Test Contract Format}, label=lst:output_test_contract_format]
 // SPDX-License-Identifier: UNLICENSED
 pragma solidity {pragma}; 
 pragma abicoder v2;
 import {"{Test, console2}"} from "forge-std/Test.sol";
 import "../src/{ contractname+".sol" }";
 contract {contractname}Test is Test {
     // Body of Test contract
 }
\end{lstlisting}

\paragraph{Prompt Chaining} In this method~\cite{b15}, we have divided the process of property generation into 3 different sub-tasks. I) Understanding each function of the smart contract and generating a list of functionalities, as shown in Listing \ref{lst:PC-Prompt1}. II) Generating properties for each functionality in natural language (English), as shown in Listing \ref{lst:PC-Prompt2}. III) Converting properties in English into foundry test properties, in solidity. These three prompts were chained together to generate the final output. The same zero-shot prompt’s output format was used here as well.
It was repeatedly observed in the output of the zero-shot prompt that LLM provides comments on how to write properties instead of giving the actual code. In this regard, our methodology got inspiration from~\cite{hassan2024llm} in using a two-step approach, 1) list down all the properties in English, 2) convert the listed properties to Solidity. Generating all the properties in English first, resolved the issue. This method is more suitable for complex tasks, as LLM works on each subtask sequentially and gives a more detailed answer. In addition to achieving better results, prompt chaining provides ease in debugging the problems in the responses of LLMs.

% Listings 2,3

\begin{lstlisting}[style=englishStyle, caption={Prompt Chaining Prompt-I}, label=lst:PC-Prompt1]
Your task is to explain all the functionalities of the given smart contract in detail.
Here is the input smart contract:
{smartcontract}
{referencecontracts}
\end{lstlisting}

\begin{lstlisting}[style=englishStyle, caption={Prompt Chaining Prompt-II}, label=lst:PC-Prompt2]
Your task is to explain and list down all the test properties in english for the foundry test contract that can be generated from the given smart contract, in detail to cover each edge case.

Here is the input smart contract:
{smartcontract}
{referencecontracts}

The following is the explanation of the functionality of the given smart contract so you can better understand the given smart contract:{functionalities}
\end{lstlisting}

\paragraph{Few-shot prompt} In the few-shot prompts~\cite{b16}, we retrieve two example smart contracts, similar to the input smart contract. Each example includes a smart contract and its corresponding test properties. We pass the retrieved examples along with the input smart contract and any imported contracts it references. To retrieve the examples, we first convert the source smart contract from the input data into vector embeddings using OpenAI’s \emph{text-embedding-ada-002} embeddings model. We have already stored the vector embeddings of the example dataset locally in Chroma Db. We then compare these embeddings to the embeddings of the input data using cosine similarity to find the two most similar examples. Instead of defining a fixed similarity threshold, we employ a KNN search and retrieve the two nearest neighbors as examples. These examples, along with the input smart contract, are all selected from the Solady utility libraries to ensure relevance. However, we exclude the input smart contract itself and its test smart contract from the retrieved examples.  

Since we retrieve two examples, we refer to this as a 2-shot prompt in the following sections. While increasing the number of examples could potentially provide more context, it also introduces a higher likelihood of retrieving irrelevant examples. To maintain relevance and prevent dilution of useful information, we limited the retrieval to only two examples.

We did not enforce a minimum similarity threshold when selecting examples. This decision ensures that even if the input smart contract does not have closely related contracts in the dataset, we can still provide some context to the LLM. By doing so, the model receives relevant structural and semantic cues, allowing it to generate meaningful test properties even in cases where highly similar examples are unavailable.

Both the zero-shot and prompt chaining approaches generated executable properties. However, the outputs still had multiple issues, which we discuss in Section~\ref{sec:eval}. Providing similar examples as context to the LLM appeared to be a suitable approach to address these issues, as it offers the LLM some context to understand the nature of the expected output. Since both the input smart contract and the retrieved examples come from Solady, we can use Solady's test properties as a benchmark for evaluating the generated properties. We aimed to retrieve examples from the same domain, if available in the dataset, or at least examples with similar semantics to the input smart contract. To achieve this, we employed the cosine similarity approach with the KNN-based selection of two nearest examples.

In the next section, we discuss the experimental evaluation to showcase the effectiveness of our LLM-assisted property generation approach for Solidity-based smart contracts.

% Evaluation Table
\begin{table*}[ht]
\renewcommand{\arraystretch}{1.3} % Increase row spacing
    \centering
    \begin{threeparttable}
        \caption{Evaluation of LLMs against prompting techniques for generating properties for Solidity smart contracts}
    \label{table:results}
        \begin{tabular}{|l|l|c|c|c|c|c|c|}
            \hline
            \textbf{LLMs} & \textbf{Prompts} & \textbf{PG (AVG)} & \textbf{EP (\%)} & \textbf{NEP (\%)} & \textbf{OAEP (\%)} & \textbf{Mutation Score (\%)} & \textbf{Inference time (s)} \\ \hline
            \multirow{3}{*}{GPT-4o} & Zero-Shot & 17.625 & 85.92 & 14.08 & 41.66 & 16.639 & 34.24 \\ \cline{2-8}
            & Prompt Chaining & 21.12 & 81.05 & 18.95 & 30.76 & 20.812 & 57.69 \\ \cline{2-8}
            & 2-Shot & 12.95 & 88.01 & 11.99 & 47.82 & 18.769 & 27.52 \\ \hline
            \multirow{3}{*}{GPT-4-Turbo}& Zero-Shot & 11.04 & 75.79 & 24.21 & 25 & 15.55 & 37.23 \\ \cline{2-8}
            & Prompt Chaining & 12.63 & 79.21 & 20.79 & 29.16 & 18.943 & 105.6 \\ \cline{2-8}
            & 2-Shot & 10 & 74.92 & 25.08 & 32.25 & 8.418 & 45.2 \\ \hline
            \multirow{3}{*}{Gemini-1.5-pro}& Zero-Shot & 22.64 & 75.78 & 24.22 & 18.18 & 15.54 & 40.06 \\ \cline{2-8}
            & Prompt Chaining & 28.68 & 89.62 & 10.38 & 31.57 & 25.997 & 114.57 \\ \cline{2-8}
            & 2-Shot & 25.35 & 78.61 & 21.39 & 31.57 & 8.868 & 59.71 \\ \hline 
            \multicolumn{6}{|c|}{Human written test properties by Solady} & 31.746 & \\ \hline
        \end{tabular}
        \begin{tablenotes}
            \footnotesize
            \item \textbf{NOTE:} PG: Properties Generated, EP: Executable Properties, NEP: Non-Executable Properties, OAEP: Outputs with All Executable Properties
            \item * The inference time for prompt chaining is the sum of the inference times for all three calls.
            \item * All figures represent average values for the smart contracts.

        \end{tablenotes}
    \end{threeparttable}
\end{table*}

% -----------------------

\section{Evaluation}\label{sec:eval}
We evaluated LLMs for their ability to generate test properties against two primary questions: can LLMs generate effective test properties? If so, which prompting technique is the most effective for which LLM? %To address these questions, we assessed the effectiveness of outputs from different LLMs using various prompting techniques.

\subsection{Test Setup}\label{AA}
We used three LLMs: OpenAI's \textit{GPT-4o}, \textit{GPT-4Turbo}, and Google AI's \textit{Gemini-1.5-pro}. We accessed all three LLMs through their respective APIs: \texttt{gpt-4o}, \texttt{gpt-4-turbo-2024-04-09}, and \texttt{gemini-1.5-pro-latest}. We set the parameter \texttt{temperature=0} for all LLMs.  Additionally, we used \texttt{max\_tokens=4096} for GPT-4Turbo. All experiments were conducted on a laptop running Ubuntu 22.04 LTS OS, equipped with 16 GB of RAM and a 12th Gen Intel(R) Core(TM) i5-1235U 1.30 GHz processor. We used \textit{Langchain} (Python version)\cite{langchain} for developing the pipeline. The version of Forge (Foundry Compiler) used is \texttt{forge 0.2.0}. We utilized \textit{Sumo}\cite{b13} for automated mutation testing. We used this tool because it features 25 Solidity-specific mutation operators, as well as 19 traditional operators, some of them are listed in Table \ref{mutation operators}. For example, one mutation changes the += operator to -=, producing a mutant that is syntactically correct but semantically different, thereby helping to evaluate the effectiveness of test properties in detecting subtle behavioral changes.
% \texttt{max\_tokens=4096}. For the Gemini model, the following parameters were used: \texttt{temperature=0}, \texttt{request\_timeout=3600}.

\subsection{Dataset}
We used human-written test properties provided by Solady~\cite{b9} as examples in few-shot prompts and as a human baseline to evaluate the performance of LLMs.

Solady is an open-source repository featuring highly optimized Solidity code snippets of various categories, including accounts, auth, tokens, and utils. It is known for its gas optimization techniques and efficiency enhancements~\cite{b45}. It provides efficient implementations of commonly used libraries, such as MerkleProofLib, along with advanced features like LibZip. By integrating low-level inline assembly within user-friendly APIs, Solady streamlines the development of clean and efficient Solidity code. Beyond being a library, it also serves as a learning platform and an experimental hub for exploring innovative gas-optimization techniques~\cite{b44}.

For this research, we focused solely on the \verb|utils| smart contracts, up to commit~\cite{b9}. The Solady utility libraries (utils) consist of 35 smart contracts that are commonly used across multiple smart contracts, making them a critical component of blockchain development. A vulnerability in such libraries can pose significant risks, as a single bug can propagate across numerous contracts, potentially leading to widespread security issues.

The Solady dataset provides a Foundry test contract for each smart contract. We perform mutation testing on these test contracts and use their average mutation score as a metric in our research, as explained in the next section.

\subsection{Evaluation Metrics}
%Table~\ref{table:results} presents the evaluation results for all metrics used to assess the properties generated by LLMs for Solidity smart contracts. 
We benchmark the quality of properties generated by the LLMs using the mutation score, which is the ratio of the number of killed mutants to the total number of non-equivalent mutants. 
The higher the mutation score, the more effective the property is at catching bugs. We compare these mutation scores with the baseline mutation score achieved by human-written properties.
%Moreover, we measure the effectiveness and efficiency of different prompting techniques in generating properties across various LLMs by the percentage of executable properties (EP), non-executable properties (NEP) and outputs with all executable properties (OAEP), and the inference time.
Moreover, we measure the effectiveness of different prompting techniques in generating properties across various LLMs by three metrics: the percentage of executable properties (EP), the percentage of non-executable properties (NEP), and the percentage of LLM outputs that have all executable properties (OAEP).
To measure the efficiency of prompting techniques, we use inference time.

Results using these metrics are presented in Table~\ref{table:results}. Please note that each metric in the table represents the average value across all the tested smart contracts.
Mutation scores for individual contracts are detailed in Table~\ref{table:mt_results}. Below we explain each metric in detail.

\paragraph{Properties Generated (PG)} %All three evaluated LLMs have output context windows of different sizes, which can affects the number of properties they generate. 
%Due to various reasons, LLMs respond to the same prompt differently. 
LLMs are inherently non-deterministic resulting in varying responses to the same prompt.
%In our scenario, this can affect the number of properties they generate.
Consequently, this variability extends to the number of properties they generate.
%To evaluate this impact, we define the \text{``Properties Generated''} metric as the average number of properties generated by an LLM across all tested smart contracts.
To evaluate this non-determinism, we define the \text{``Properties Generated''} metric as the average number of properties generated by an LLM across all tested smart contracts.

\paragraph{Executable Properties (EP \%)} This metric defines the percentage of executable properties generated out of the total properties generated by an LLM. Executable properties are those that are free from compilation errors, fault-free, and usable for testing a smart contract.

\paragraph{Non-Executable Properties (NEP \%)} The main objective of generating test properties through LLMs is to reduce manual effort.
%Therefore, generated properties should not require significant manual intervention to be considered of high quality.
Hence, generated properties should not require significant manual intervention.
The NEP\% metric defines the percentage of faulty properties that require significant manual effort to be effective, either due to compilation issues or logical errors. Compilation issues prevent the code from being executable, while logical issues necessitate human intervention to understand and rectify the LLM-generated property.

\paragraph{Outputs With All Executable Properties (OAEP\%)} This metric defines the percentage of outputs generated by LLMs in which all the generated properties are executable. In other words, it is the percentage of generated outputs having EP score as 100\%.

\paragraph{Mutation Score \%} Mutation testing is a method used to evaluate the quality of test inputs by introducing small changes (mutants) to the original code and assessing whether the test inputs can detect (kill) these changes. 
In our case, the objective is to check if test \emph{properties} can kill mutants. %, as the test inputs are automatically generated by fuzzing. 
The mutation score measures the quality of generated properties as the ratio of the number of mutants killed to the total number of non-equivalent mutants generated.
A higher mutation score indicates test properties of higher quality because they can detect more mutants.
For mutant generation and mutation testing, we used Sumo~\cite{b13}, a tool that facilitates automated mutation testing for Foundry.
We first performed mutation testing on human-written properties provided by Solady and used their score as the human baseline.
Then we tried to kill the same mutants using LLM-generated properties. 

\paragraph{Inference Time} Inference time is the average duration, in seconds, an LLM takes to generate an output given a prompt. In the case of prompt chaining, the inference time is the sum of individual inference times for all three API calls.

\subsection{Case Study: LibBit Smart Contract}

To evaluate the effectiveness of our approach, we selected the \texttt{LibBit} smart contract from the Solady repository. This contract provides a collection of utility functions for bit manipulation and boolean operations. We showcase a partial version of this library with three representative functions: \texttt{fls}, \texttt{reverseBytes}, and \texttt{or}.

\begin{lstlisting}[language=Solidity, caption={Partial implementation of LibBit.sol}, label={lst:libbit}, basicstyle=\ttfamily\footnotesize, breaklines=true,]
library LibBit {
    function fls(uint256 x) internal pure returns (uint256 r) {
        assembly {
            r := or(shl(8, iszero(x)), shl(7, lt(0xffffffffffffffffffffffffffffffff, x)))
            r := or(r, shl(6, lt(0xffffffffffffffff, shr(r, x))))
            r := or(r, shl(5, lt(0xffffffff, shr(r, x))))
            r := or(r, shl(4, lt(0xffff, shr(r, x))))
            r := or(r, shl(3, lt(0xff, shr(r, x))))
            r := or(r, byte(and(0x1f, shr(shr(r, x), 0x8421084210842108cc6318c6db6d54be)),
            0x07060605060205040602030205040301
            06050205030304010505030400000000))
        }
    }

    function reverseBytes(uint256 x) internal pure returns (uint256 r) {
        unchecked {
            uint256 m0 = 0x100000000000000000000000000000001 * (~toUint(x == uint256(0)) >> 192);
            uint256 m1 = m0 ^ (m0 << 32);
            uint256 m2 = m1 ^ (m1 << 16);
            uint256 m3 = m2 ^ (m2 << 8);
            r = (m3 & (x >> 8)) | ((m3 & x) << 8);
            r = (m2 & (r >> 16)) | ((m2 & r) << 16);
            r = (m1 & (r >> 32)) | ((m1 & r) << 32);
            r = (m0 & (r >> 64)) | ((m0 & r) << 64);
            r = (r >> 128) | (r << 128);
        }
    }

    function or(bool x, bool y) internal pure returns (bool z) {
        assembly {
            z := or(iszero(iszero(x)), iszero(iszero(y)))
        }
    }
}
\end{lstlisting}

We used Gemini (zero-shot) to generate properties for testing the functions in \texttt{LibBit}. Below is a partial snapshot of the test suite generated.

\begin{lstlisting}[language=Solidity, caption={Partially generated test cases for LibBit}, label={lst:libbittest}, basicstyle=\ttfamily\footnotesize, breaklines=true]
contract LibBitTest is Test {
    // Non-executable due to incorrect logic
    // function testReverseBytes() public {
    //     assertEq(LibBit.reverseBytes(0), 0);
    //     assertEq(LibBit.reverseBytes(1), uint256(1) << 248);
    //     assertEq(LibBit.reverseBytes(0x0102030405060708090a0b0c0d0e0f10),0x100f0e0d0c0b0a090807060504030201);
    //     assertEq(LibBit.reverseBytes(type(uint256).max), type(uint256).max);
    // }

    // Non-executable due to incorrect assertion
    // function testFuzzFls(uint256 x) public {
    //     uint256 result = LibBit.fls(x);
    //     if (x == 0) {
    //         assertEq(result, 256);
    //     } else {
    //         assertGe(result, 0);
    //         assertLt(result, 256);
    //         assertTrue((x >> result) == 0);
    //     }
    // }

    function testFls() public {
        assertEq(LibBit.fls(0), 256);
        assertEq(LibBit.fls(1), 0);
        assertEq(LibBit.fls(2), 1);
        assertEq(LibBit.fls(3), 1);
        assertEq(LibBit.fls(uint256(1) << 255), 255);
    }

    function testFuzzReverseBytes(uint256 x) public {
        uint256 result = LibBit.reverseBytes(x);
        if (x == 0) {
            assertEq(result, 0);
        } else {
            assertEq(result, LibBit.reverseBytes(x));
        }
    }

    function testOr() public {
        assertTrue(LibBit.or(true, true));
        assertTrue(LibBit.or(true, false));
        assertTrue(LibBit.or(false, true));
        assertFalse(LibBit.or(false, false));
    }
}
\end{lstlisting}

We categorize tests as \textit{non-executable properties} when they contain incorrect logic, as seen in the commented-out functions. For instance, in \texttt{testReverseBytes}, the third assertion fails because the expected value is incorrectly computed. Similarly, the final assertion in \texttt{testFuzzFls} checks whether the shifted result is 0, whereas it should be checking for 1.

Despite these non-executable properties, the remaining generated properties are logically sound and pass all tests. We then used a mutation testing tool to generate mutants of the original \texttt{LibBit} contract and executed the passing properties on these mutants.

\subsubsection*{Mutation Testing Results}

The mutation score is calculated as follows:

\[
\text{Mutation Score} = \frac{\text{Number of Mutants Killed}}{\text{Total Number of Mutants}} \times 100
\]

The properties generated by Gemini (zero-shot) achieved a mutation score of \textbf{69.74\%}, which is close to the score obtained using the human-written tests available in the official Solady repository, which achieved a score of \textbf{74.83\%}. This demonstrates that even with zero-shot generation, the model can produce tests that are effective in detecting mutants and verifying contract correctness.

\subsection{Effectiveness of LLM-generated properties}
During our experimentation, LLMs demonstrated impressive capabilities to test smart contracts. We can explain the effectiveness of the LLM-generated properties through the following two points.
\begin{itemize}
    \item Quality of properties generated by LLMs
    \item Quantity of properties generated by LLMs
\end{itemize}

\begin{figure*}
      \centering
      \includegraphics[width=1\linewidth]{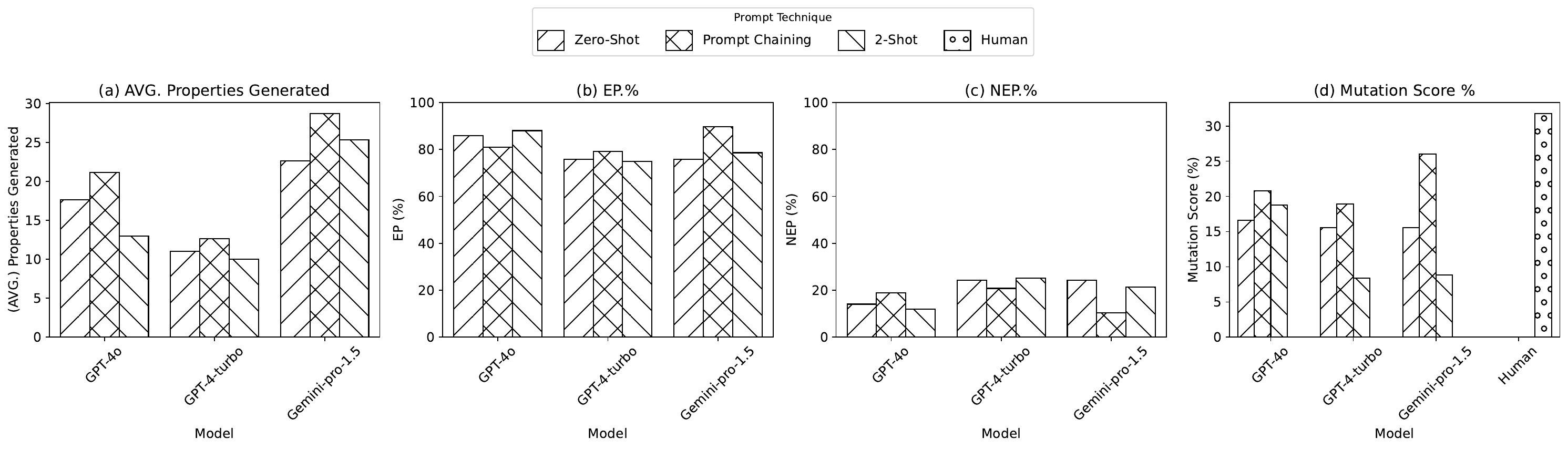}
      \caption{Bar charts (a) - (c) show the effectiveness of prompting techniques for generating properties on LLMs using PG, EP and NEP mertics, respectively. (d) shows the quality of LLM-generated properties using the Mutation Score metric.}
      \label{Bargraphs}
\end{figure*}

\paragraph{Quality of properties generated by LLMs} For this research, we consider properties to be effective if their mutation score is close to the mutation score achieved by human-written properties.
The mutation score defines the percentage of mutants (artificially induced vulnerabilities) that the test properties can detect.
The human-written properties in Solady test contracts achieved an average mutation score of \textbf{31.7\%}. Gemini 1.5 Pro achieved \textbf{25.99\%}, \textbf{15.54\%}, and\textbf{ 8.8\%} on prompt chaining, zero-shot, and few-shot prompts, respectively. In contrast, GPT-4o achieved \textbf{20.8\%}, \textbf{16.6\%}, and \textbf{18.7\%}, and GPT-4Turbo achieved \textbf{18.9\%}, \textbf{15.5\%}, and \textbf{8.4\%} on the same prompts, respectively.
Fig.~\ref{Bargraphs}d visualizes mutation scores on LLM-generated properties.

The results of prompt chaining identify LLMs as good step-by-step reasoners for smart contract testing.
However, other prompting approaches also highlight the limitations of LLMs for this use case.
The mutation scores on zero-shot and 2-shot prompts are \textbf{16.64\%} and \textbf{18.77\%} for GPT-4o, \textbf{15.55\%} and \textbf{8.42\%} for GPT-4Turbo, and \textbf{15.54\%} and \textbf{8.86\%} for Gemini-pro-1.5. 

Few-shot prompts scored the lowest among all. A few-shot prompt depends on retrieving relevant examples, and a possible reason for the low score could be the retrieved examples. In most cases, the retrieved examples were not from the same domain as the input smart contract, due to a small dataset. This would highlight that context-specific retrieval is crucial for good performance in this use case. An ideal dataset for few-shot would contain smart contracts from every domain, leading to a large dataset

\paragraph{Quantity of properties generated by LLMs} 
The bar graphs in Fig.~\ref{Bargraphs}a, Fig.~\ref{Bargraphs}b and Fig.~\ref{Bargraphs}c show the average number of generated properties, the average percentage of executable properties and the average percentage of non-executable properties, respectively.
They indicate that the quantity of generated properties depends on both 
%the size of 
the LLM %'s output context window 
and the prompting technique used.
On average, Gemini-pro-1.5 generated more properties per contract. %It offers an output context window of 8192 tokens, whereas OpenAI’s GPT-4 Turbo and GPT-4o offer output context windows of 4096 tokens and 2048 tokens, respectively.

%Ideally, a high number of generated properties alone should not be sufficient to consider the output good. We would want a high ratio of executable and useful properties in the output by arguing that the more executable properties we have, the more chances we have to detect vulnerabilities. However, Table~\ref{table:results} hints at a positive correlation between the average mutation score and the average number of generated properties. 

%However, a high number of generated properties alone is not sufficient to consider the output good. 
%The output should contain a high ratio of executable and useful properties. 
%This is important for effectiveness because the more executable properties we have, the more chances we have to detect vulnerabilities.
%As evident from the results in Table~\ref{table:results}, an increase in the average mutation score was observed whenever there was an increase in the average number of generated properties. 

We observed a high variation in both the quantity and the ratio of executable properties with different prompting techniques.
%Prompt chaining not only produced more properties but also had a higher ratio of executable properties compared to other prompting techniques for the respective LLM.
Though prompt chaining consistently produced more properties across all LLMs, it underperformed with respect to executable properties on GPT-4o.
With prompt chaining, Gemini-pro-1.5 generated \textbf{28.68} properties on average with \textbf{89.62\%} executable properties (EP).
These are very significant numbers considering the performance of other LLMs and even the performance of Gemini-pro-1.5 on other prompting techniques. 
%Overall, prompt chaining resulted in a higher number of generated properties on all LLMs, considering \textbf{21.12} with EP=\textbf{81.05\%} and \textbf{12.63} with EP=\textbf{79.21\%} for GPT-4o and GPT-4 Turbo, respectively (Fig.~\ref{Bargraphs}a and Fig.~\ref{Bargraphs}b).

\subsection{Limitations  of LLMs}\label{SCM}
Multiple compilation issues were commonly observed in the outputs of all prompts. Most were manually resolvable without much effort, making the contracts executable. However, some others required significant human effort to consider the generated properties as executable. Both cases are explained in the next sub-sections.

\subsubsection{Common Issues Requiring High Human Effort:}

The following issues require substantial manual effort to resolve, which contradicts the main objective of this research. Therefore, outputs that contain these issues were ignored and not considered for the mutation testing phase. 

\begin{itemize}
    % \item \textbf{Comments Instead of Code:} Some test properties were left unimplemented, with only comments added instead of code. Sometimes, comments mentioned possible properties for the provided smart contract. This issue was mostly observed with zero-shot prompts, prompting us to include a list of possible properties in the prompt.
    \item     \textbf{Bogus Property:} In some of the properties logical errors were observed, such properties are referred to as bogus properties. For example, in the Listing~\ref{lst:bgp}, an example from MinHeapLib smart contract is mentioned. A value variable was compared with the heap root using a \verb|less than| operator, followed by an equality comparison in an assertEq statement.
% Listings 1

\begin{lstlisting}[style=solidityStyle, caption={Example of Bogus Logic}, label=lst:bgp]
// Logical error
if (value < heap.root()) {
    assertEq(heap.root(), value);
}
\end{lstlisting}

    \item \textbf{Test Contract Pointing to Irrelevant Functions:} In some cases, test contracts called functions not implemented in the source contract or imported libraries, causing compilation issues. It does not make sense to implement these functions manually. This issue was observed with all prompting techniques used.
% --------------------------
\end{itemize}

\subsubsection{Common Issues Requiring Minor Human Effort}

These issues were manually resolvable with minimal effort, and the test contracts considered for mutation testing included these issues.

\begin{itemize}
    \item \textbf{Creating Objects of Libraries:} Solidity does not support creating objects of libraries. However, in some outputs, an object of a library was created inside the setup function, causing a compilation error.
    \item \textbf{Inserting Emojis Instead of Unicode:} In the LibString and Base64 smart contracts, LLMs initialized a bytes-type variable with emojis instead of their Unicode equivalents, causing compilation errors.

\begin{lstlisting}[style=englishStyle, caption={Error: emojis instead of unicode}, label={PC-Prompt1}]
Compiler run failed:
Error (8936): Invalid character in string. If you are trying to use Unicode characters, use a unicode"..." string literal.
--> test/Base64.t.sol:43:35:
    
bytes memory data = bytes("emojis here");
\end{lstlisting}
    
    \item \textbf{Incorrect Epoch Day Conversion:} In the DateTimeLib test contract, the function converting a date to an epoch day resulted in incorrect values. For example, \verb|2023-03-15| should equal \verb|19426| but resulted in \verb|19431|. This type of issue occurred in multiple instances.

\begin{lstlisting}[style=englishStyle, caption={Error: incorrect epoch day conversion}, label={PC-Prompt1}]
"[FAIL. Reason: 2023-03-15 should be epoch day 19426: 19431 != 19426]" 
    
testDateToEpochDay() (gas: 4803)
\end{lstlisting}

    \item \textbf{Creating Mocks Inside the Contract Body:} Solidity does not support the dynamic creation of mock contracts or dynamic modification of contract behavior at runtime.

    \item \textbf{Address with invalid checksum:} Address values assigned to address variables had invalid checksums. Ethereum addresses use checksums to ensure validity. Solidity requires addresses in a check-summed format, involving specific letter capitalization.

\begin{lstlisting}[style=englishStyle, caption={Error: invalid checksum}, label={PC-Prompt1}]
Compiler run failed:
SyntaxError:
This looks like an address but has an invalid checksum. Correct checksummed address: 0x1234aBCD1234ABCD1234ABCD1234ABCD1234. If this is not used as an address, please prepend '00'. For more information please see docs.soliditylang.org/en/develop/types.html#address-literals
test/LibString.t.sol:32:29:
|
32 | address testValue = 0x1234ABCD1234ABCD1234ABCD1234ABCD1234;
}
\end{lstlisting}

\end{itemize}

\subsection{Observations}
\label{subsec:observations}
The following are a few behaviors observed during the experimentation. 

\begin{itemize}
    \item \textbf{High rate of hallucinations:} We observed a high rate of variation in the number of properties generated for a specific smart contract on different runs. For instance, in the first run, the number of generated properties might be 15, but in the second run, it could drop to 5. This issue was mostly observed with the zero-shot prompts. However, prompt chaining and the few-shot approach significantly reduced the hallucinations. The difference in the count of output properties reduced to 1 or rarely 2 properties.
    \item \textbf{LLMs as compiling agents:} We developed an agent for resolving compilation issues using Langchain and integrated it with the Foundry compiler. Whenever a compilation error occurs, the agent adopts a reasoning and action approach to correct it. We evaluated GPT-3.5-Turbo and GPT-4-Turbo as agents for resolving these issues. It is important to note that this experimentation was conducted using the versions of the LLMs available up until December 2023. Both OpenAI models demonstrated good capabilities in resolving compilation issues. GPT-4-Turbo showed superior reasoning compared to the GPT-3.5-Turbo. However, GPT-3.5-Turbo provided faster reasoning and action. Despite these capabilities, the compiling agent could only resolve issues that did not require significant human effort. Consequently, we performed this task of resolving errors manually instead of making API calls. Although we did not explore the agentic approach extensively in this research, it could be considered for future work.
    \item \textbf{Experience with zero-shot prompts:} Initially, we tried the zero-shot prompt without any output format. However, we repeatedly observed that the generated test contract did not contain all the necessary imports for the Foundry test contract, causing many compilation issues. To avoid these compilation issues, we made it mandatory for the LLM to follow the given structure for the output.

     Some test properties were left unimplemented, with only comments added instead of code. Sometimes, comments mentioned possible properties for the provided smart contract.
\end{itemize}

\section{Discussion}
We measured the quality of generated properties using mutation testing. This approach necessitates understanding the specific types of mutations applied to the Solidity code. We used a total of 44 different mutation operators: 19 traditional mutation operators and 25 Solidity-specific operators. These mutation operators were automatically applied to the smart contracts to generate mutants using Sumo~\cite{b13}. Table~\ref{mutation operators} lists some of the mutation operators used.
% The first 19 operators, from \verb|ACM| to \verb|UORD|, are traditional mutation operators, and the last 25 operators, from \verb | AVR| to \verb | VVR|, are solidarity-specific operators.

In Section~\ref{sec:eval}, we have discussed the average mutation scores. 
We mentioned the average score because some contracts produced executable outputs with one LLM or prompt technique but not with others.
Table~\ref{table:mt_results} provides a complete overview of mutation scores in all settings for each smart contract. 
Cells containing `N/E' indicate that the generated properties were not executable, even after trying 3 to 5 different calls.
Notably, each contract's output was generated a minimum of 3 times and a maximum of 5 times if we did not get an executable output.

% In Section~\ref{sec:eval}, a metric that was not discussed in detail before is OAEP\%.
In terms of generating a higher percentage of those outputs in which all properties were executable, GPT-4o leads other LLMs by a significant margin, with 47.82\% OAEP on 2-shot prompts and 41.66\% on zero-shot prompts. 
If we evaluate prompting techniques based on this metric, Table~\ref{table:results} shows that 2-shot prompting achieved a higher OAEP score compared to other prompting techniques. 
However, it did not increase the mutation score for 2-shot prompting. 
Therefore, an increased ratio of executable properties alone is not sufficient for vulnerability detection.

Table~\ref{table:mt_results} provides mutation scores for all smart contracts tested during experimentation. 
On some smart contracts, all settings demonstrated almost similar behaviour.
For example, on \verb|LibBit|, all LLMs achieved high mutation scores on both zero-shot and prompt chaining.
However, there were other cases like \verb|CREATE3|,  \verb|Multicallable|, \verb|MerkleProofLib| and \verb|DeploylessPredeployQueryer|, where all LLMs on all prompting techniques either generated a non-executable output or achieved zero mutation score. 
On other smart contracts such as \verb | LibClone | and \verb | ERC1967Factory |, a very low mutation score was observed in all settings.
The mutation score of \verb|ERC1967FactoryConstants| is not available because Solady does not provide any tests for it.   

\section{Related work}
 Smart contracts are still prone to many loopholes that can be exploited by attackers despite significant advancements in blockchain technology and the rigorous auditing processes they undergo.
 In a recent study~\cite{b17}, blockchain exploitations are classified into application bugs, protocol bugs, cryptography flaws, and implementation errors. 
 Although these classifications are broad, our research explores application bugs, which usually occur due to vulnerable smart contracts. 
 A few studies~\cite{b18},~\cite{b19} have shown that these application bugs, such as integer overflow~\cite{b1}, 
 re-entrancy~\cite{b2}, front-running~\cite{b22}, and access control~\cite{b23} vulnerabilities, can lead to financial losses of millions of dollars, as evidenced by the Parity wallet attack, where an attacker gained access to a wallet and stole 150,000 ETH, the DAO attack, and other high-profile incidents.

To mitigate these smart contract vulnerabilities, various approaches have been developed to enhance the overall security posture of smart contracts, including static analysis tools~\cite{b24},~\cite{b25},~\cite{b26} and dynamic analysis tools~\cite{b27},~\cite{b28},~\cite{b29}.
Static analysis tools use data-flow analysis to find vulnerable patterns and require users to provide customized specifications.
These tools include Oyente~\cite{b29}, Manticore~\cite{b27}, Mythril~\cite{b28}, and MadMax~\cite{b25}.
On the other hand, dynamic analysis tools use symbolic execution to explore the state of the program, identifying vulnerabilities through simulated execution paths.
Formal verification has also been widely adopted to ensure the correctness of smart contracts~\cite{b30},~\cite{b31}.
It rigorously proves the correctness of contracts against formal specifications, but it can be complex and time-consuming to implement.

Fuzzing has recently emerged as a successful technique for identifying potential vulnerabilities in traditional programs~\cite{b32},~\cite{b33},~\cite{b34}.
Due to its dynamic nature, it can discover unexpected bugs without the overhead of predefined vulnerable patterns.
In recent years, various fuzzers~\cite{b35},~\cite{b36},~\cite{b37} have been utilized to uncover smart contract vulnerabilities.

Various studies have employed both static~\cite{b1},~\cite{b40} and dynamic inferences in the property generation of smart contracts. 
For instance, SolType~\cite{b1} uses static inference techniques but is limited to properties that cannot extend beyond arithmetic operations, leaving room for integer overflow and underflow vulnerabilities in smart contracts.
In contrast, InvoCon~\cite{b38} used static inference to detect invariants but the invariants remained unverified.
However, its advanced version, InvoCon+~\cite{b39}, generates smart contract invariants using both static and dynamic inference but requires contracts to have more transaction histories.

Recently, LLMs have been employed in detecting smart contract vulnerabilities. 
Various tools like TitianFuzz~\cite{b41} and FuzzGpt~\cite{b42} use LLM-guided fuzzing for deep learning libraries. 
LLMs are also extensively used in several program repair tasks.
Our work aligns with previous LLM-based property generation techniques for smart contracts, such as PropertyGPT~\cite{b6} and SmartInv~\cite{b43}.
Both studies have generated extensive high-quality properties using LLMs.
LLMs have also been explored in hardware verification. For example, LASP~\cite{ayalasomayajula2024lasp}  presents a framework that combines static RTL analysis with LLMs to generate security properties for SoC verification. The generated natural language properties are easily translatable into SystemVerilog Assertions (SVAs), with an 80 percent success rate across benchmarks like RSA and AES. Although focused on hardware, LASP shares core goals with our work, such as using LLMs to automate property generation from structured code, highlighting the broader applicability of LLMs in formal verification.

Beyond smart contracts, LLMs are also being explored for generating property based tests in other complex and critical domains. For instance, ~\cite{etemadi2025llm} introduced an LLM based approach for generating property based tests to guardrail Cyber Physical Systems (CPSs). Their tool, CHEKPROP, uses a two phase approach: property based test generation at design time and monitoring at runtime. This demonstrates the broader applicability of LLMs in ensuring system safety through automated property testing. Although applied to a different domain, their work emphasizes the growing trend and effectiveness of using LLMs for automated property extraction and test generation, reinforcing the methodology and findings of our study.
Recently, deep learning techniques have been applied to smart contract verification. Cider~\cite{b46}, a reinforcement learning-based tool, infers contract invariants to ensure arithmetic safety. By formulating invariant generation as a Markov Decision Process (MDP), it learns a neural policy to predict useful invariants, improving verification efficiency. Unlike our work, which guides LLMs for property generation, Cider focuses on reinforcement learning for invariant inference. Our research distinguishes itself by evaluating different state-of-the-art LLMs on different prompting techniques and guiding them to generate high-quality smart contract properties.
\begin{table*}[h]
	\centering
	\caption{Mutation Score (\%) achieved on individual \textbackslash Solady\textbackslash utils smart contracts.}
	\label{table:mt_results}
	\begin{adjustbox}{max width=\textwidth}
		\renewcommand{\arraystretch}{1.5} % Increase row spacing
		\begin{tabular}{|c|c|c|c|c|c|c|c|c|c|c|}
			\hline
			\multirow{2}{*}{\textbf{Smart Contract}} & \multirow{2}{*}{\textbf{Human}} & \multicolumn{3}{c|}{\textbf{Zero Shot}} & \multicolumn{3}{c|}{\textbf{Prompt Chaining}} & \multicolumn{3}{c|}{\textbf{2-Shot}} \\ \cline{3-11} 
			&  & \textbf{GPT-4o} & \textbf{GPT-4Turbo} & \textbf{Gemini} & \textbf{GPT-4o} & \textbf{GPT-4Turbo} & \textbf{Gemini} & \textbf{GPT-4o} & \textbf{GPT-4Turbo} & \textbf{Gemini} \\ \hline
			SafeCastLib & 46.34 & N/E & 11.26 & N/E& N/E& 6.31& N/E& 12.14 & 13.69& N/E\\ \hline
			UUPSUpgradeable & 50 & N/E & 45.45 & N/E& N/E& N/E& N/E& N/E& 0& N/E\\ \hline
			LibRLP & 0 & N/E & N/E & N/E& 0 & N/E& 0 & N/E& 0 & N/E\\ \hline
			DateTimeLib & 76.59 & 66.34 & 30.26 & 47.88 & N/E& 26.05& 57.99& 31.73& 14.79& N/E\\ \hline
			DynamicBuffer & 61.36 & 1.21 & 1.15 & 2.33& 2.13& 1.72& 1.16& 7.32& 1.37& 4.56\\ \hline
			LibBit & 74.83 & 66.45 & 66.67 & 69.74 & 73.38& 74.34& 74.34& 72.9 & 0 & 0 \\ \hline
			JSONParserLib & 69.81 & 46.08 & 39.64 & N/E& N/E& 51.38& 52.68& 37.46& 38.35& 0 \\ \hline
			LibBitMap & 80.6 & N/E & 25.85 & 19.24& 50.36& 28.43& 28.19& N/E& 20.74& 27.99\\ \hline
			RedBlackTreeLib & 51.62 & N/E & N/E & N/E& N/E& N/E& N/E& 11.14& N/E& N/E\\ \hline
			Base64 & 25 & 23.08 & 18.75 & 15.38 & 15.38& 30.77& 18.75& 30.77& 10.53& 30.77\\ \hline
			SSTORE2 & 33.33 & 11.11 & 11.11 & 11.11 & 10.53& 11.11 & 21.05& 11.11 & 11.11& N/E\\ \hline
			SignatureCheckerLib & 8.33 & N/E & N/E & N/E& N/E& N/E& 62.5& 10.53& 2.38& 9.52 \\ \hline
			ReentrencyGuard & 0 & 0 & N/E & N/E& N/E& N/E& N/E& 0 & N/E& N/E\\ \hline
			Create3 & 0 & 0 & 0 & 0 & 0 & 0 & N/E& 0 & 0 & N/E\\ \hline
			EIP712 & 42.55 & N/E & N/E & 6.32& 0 & N/E& N/E& 27.66& 10.64& 6.38 \\ \hline
			EnumerableSetLib & 15.19 & 16.44 & 16.44 & 16.44& 12.77& 5.83& 13.7& 10.96& 0 & 16.44\\ \hline
			Multicallable & 0 & 0 & N/E & N/E& 0 & N/E& 0 & 0 & 0 & 0 \\ \hline
			Initializable & 20 & 20 & N/E & 20& N/E& N/E& N/E& 20& 20& N/E\\ \hline
			LibString & 31.36 & N/E & 22.14 & 21.71 & 26.4& N/E& 9.38& 24.06 & 7.91& 20.53\\ \hline
			DeploylessPredeployQueryer & 0 & 0 & 0 & N/E& 0 & 0 & N/E& 0 & 0 & N/E\\ \hline
			FixedpointMathLib & 70.74 & N/E & 8.93 & 29.23 & N/E& 3& 35.93 & 26.73 & 0.11& 1.03\\ \hline
			LibZip & 20 & 0 & 0 & 0 & 20& 0 & 0 & N/E& 0 & 0 \\ \hline
			GasBurnerLib & 0 & 0 & N/E & 0 & 100 & 100& 100& N/E& 0 & N/E\\ \hline
			ERC1967Factory & 2.78 & 0 & 0 & N/E& N/E& 5.41& N/E& N/E& 0 & 0 \\ \hline
			ERC1967FactoryConstants & N/A & 42.86 & N/E & N/E& N/E& 42.86& 0 & N/E& 42.86& N/E\\ \hline
			ECDSA & 25 & N/E & 7.69 & 2.94& 28.57& 10.81& N/E& N/E& 2.7& 2.94\\ \hline
			SafeTransferLib & 0 & 2.82 & 0 & N/E& N/E& N/E& N/E& N/E& 0 & N/E\\ \hline
			MerkleProofLib & 0 & N/E & 0 & 0 & 0 & 0 & N/E& N/E& 0 & N/E\\ \hline
			MetaDataReader & 30 & 11.67 & 22.22 & 0 & 20.69& 2.5& N/E& N/E& 12& N/E\\ \hline
			LibPRNG & 62.5 & 4.48 & 1.32 & 2.63& 4.69& 0 & N/E& N/E& 0 & 0 \\ \hline
			LibClone & 6.32 & 1.48 & 0 & N/E& N/E& 1.26& N/E& N/E& 0 & 4.46\\ \hline
			LibMap & 79.03 & 31.62 & 25.85 & 19.24& 50.36& 43.77& N/E& N/E& 43.37& N/E\\ \hline
			MinHeapLib & 43.84 & N/E & 15.79 & 24.42 & 11.46& 9.09 & 13.33 & N/E& N/E& 26.14\\ \hline
			LibSort & 20.49 & 20.41 & N/E & 21.43& N/E& 0 & 4.94& 22.45 & 0 & N/E\\ \hline
			\textbf{Average} & \textbf{31.746} & \textbf{16.639} & \textbf{15.55} & \textbf{15.54} & \textbf{20.812}& \textbf{18.94}& \textbf{25.997}& \textbf{18.77} & \textbf{8.42}& \textbf{8.87}\\ \hline
		\end{tabular}
	\end{adjustbox}
	\begin{tablenotes}
		\footnotesize
		\item \textbf{NOTE:} 
		\item * Human refers to test properties by solady. 
		\item * Gemini: Gemini-1.5-pro
		\item * N/A: Not available
		\item * N/E means that the properties generated were non-executable.
		\item * 0 mutation score means, no mutants were killed by the properties.
	\end{tablenotes}
\end{table*}
\section{Conclusion}
In this paper, we proposed an approach that harnesses state-of-the-art LLMs to assist property-based testing by automatically generating high-quality properties for Solidity-based smart contracts. 
After generating the properties, we applied fuzzing to test the smart contracts and measured the quality of the generated properties through mutation testing. 
Our findings demonstrated that LLMs have the potential to generate high-quality properties comparable to those written by human experts. 
Furthermore, we extensively evaluated LLMs against various prompting techniques, including zero-shot, few-shot, and prompt chaining.
Few major findings emerged from our research. 1) dividing the task of property generation into a step-by-step approach, akin to how a human would tackle a complex problem, significantly improved the results,
%Prompting the LLM for each sub-task at a time enhanced the outcomes. 
2) using prompt chaining, all three LLMs performed better compared to other prompting techniques, with Gemini-pro-1.5 delivering the best overall results.

\begin{table}[h]
	\centering
	\caption{Examples of a Few Mutations Applied in Generating Smart Contract Mutants}
	\renewcommand{\arraystretch}{1.12} % Increase row spacing
	\begin{tabular}{|c|}
		\hline
		\textbf{Mutation Example} \\ \hline
		\texttt{+= → -= → =} \\ \hline
		\texttt{break → continue → break} \\ \hline
		\texttt{true → false} \\ \hline
		\texttt{+ → - → * → / → < → >=} \\ \hline
		\texttt{catch{} →} \\ \hline
		\texttt{if(condition) → if(false) else{} →} \\ \hline
		\texttt{enum.member1 → enum.member2} \\ \hline
		\texttt{uint256 → uint8} \\ \hline
		\texttt{hex"01" → hex"random"} \\ \hline
		\texttt{-- → ==} \\ \hline
		\texttt{1 → 0} \\ \hline
		\texttt{while(condition) → while(false)} \\ \hline
		\texttt{function overloadedF(){} →} \\ \hline
		\texttt{function f() override {} →} \\ \hline
		\texttt{x = getData() → x = super.getData()} \\ \hline
		\texttt{x = super.getData() → x = getData()} \\ \hline
		\texttt{"string" → ""} \\ \hline
		\texttt{++ → -- → ! →} \\ \hline
		\texttt{0x67ED2e5dD3d0... → address.this()} \\ \hline
		\texttt{constructor(){} →} \\ \hline
		\texttt{memory → storage} \\ \hline
		\texttt{delete →} \\ \hline
		\texttt{delegatecall() → call()} \\ \hline
		\texttt{emit Deposit(...) → /*emit Deposit(...)*/} \\ \hline
		\texttt{require(...) → /*require(...)*/} \\ \hline
		\texttt{function f() public → function f() private} \\ \hline
		\texttt{msg.value() → tx.gasprice()} \\ \hline
		\texttt{addmod → mulmod keccak256 → sha256} \\ \hline
		\texttt{function f() onlyOwner → function f()} \\ \hline
		\texttt{function f() → function f() onlyOwner} \\ \hline
		\texttt{function f() modA modB → function f() modB modA} \\ \hline
		\texttt{function f() onlyOwner → function f() onlyAdmin} \\ \hline
		\texttt{modifier m() override {} →} \\ \hline
		\texttt{function f() payable → function f()} \\ \hline
		\texttt{return amount; → //return amount;} \\ \hline
		\texttt{selfdestruct(); → //selfdestruct();} \\ \hline
		\texttt{SafeMath.add → SafeMath.sub} \\ \hline
		\texttt{msg.sender → tx.origin} \\ \hline
		\texttt{wei → ether minutes → hours} \\ \hline
		\texttt{uint private data; → uint public data;} \\ \hline
	\end{tabular}
	\label{mutation operators}
\end{table}

\section{Future work}
The following approaches can be considered for the future work,
\begin{itemize}
    \item \textbf{Combining Prompt Chaining with Fine-tuned LLM for property generation: } LLMs can be finetuned for this task of property generation. Then we can combine our most effective approach prompt chaining with the fine-tuned LLM.

    \item \textbf{Domain-specific Few-shot prompting:} We can prepare a data set containing high-quality test properties from all domains, or we can use the entire Solady dataset as an example data set and test the approach on other industry-standard smart contracts. Our goal is to provide the LLM with context that is highly related to the input smart contract.  

    \item \textbf{Agentic approach for properties generation:} Agents also follow a step-by-step reasoning and action approach to solve problems. Allowing the agent to work on each specific task one at a time may reduce errors and increase the quality of generated properties. Agents can be provided with properties in natural language and asked to code each property individually. We can also combine domain-specific examples with the agentic approach. An agent can be integrated to review the properties and correct the generated code
\end{itemize}

\printbibliography

\vspace{12pt}
\end{document}